\documentclass[pre,showpacs,tightenlines,floatfix,twocolumn,superscriptaddress]{revtex4}
\usepackage{graphicx}
\usepackage{amssymb,amsfonts,amsmath}
\usepackage[T1]{fontenc}
\usepackage{times}
\usepackage{color}
\renewcommand{\onlinecite}{\cite}

\def\gsim{\mathrel{\rlap{\lower4pt\hbox{\hskip1pt$\sim$}}\raise1pt\hbox{$>$}}}        
\def\lsim{\mathrel{\rlap{\lower4pt\hbox{\hskip1pt$\sim$}}\raise1pt\hbox{$<$}}}        

\begin{document}
\title{Re-examining the directional-ordering transition in the compass
  model with screw-periodic boundary conditions} \date{\today}
\author{Sandro \surname{Wenzel}}
\email{sandro.wenzel@epfl.ch}\affiliation{Max-Planck-Institute for
  the Physics of Complex Systems (MPIPKS), N\"othnitzer Str.\ 38, D-01187 Dresden,
  Germany} \affiliation{Institute of Theoretical Physics, \'Ecole
  Polytechnique F\'ed\'erale de Lausanne (EPFL), CH-1015 Lausanne, Switzerland}
\author{Wolfhard \surname{Janke}} \email{janke@itp.uni-leipzig.de}
\affiliation{Institut f\"ur Theoretische Physik and Centre for
  Theoretical Sciences (NTZ), Universit\"at Leipzig, Postfach 100920,
  D-04009 Leipzig, Germany} 
\author{Andreas M. \surname{L\"auchli}} \email{aml@pks.mpg.de}
\affiliation{Max-Planck-Institute for
  the Physics of Complex Systems (MPIPKS), N\"othnitzer Str.\ 38, D-01187 Dresden,
  Germany}
\begin{abstract}
  We study the directional-ordering transition in the two-dimensional
  classical and quantum compass models on the square lattice by means
  of Monte Carlo simulations. An improved algorithm is presented which
  builds on the Wolff cluster algorithm in one-dimensional subspaces
  of the configuration space. This improvement allows us to study
  classical systems up to $L=512$.  Based on the new algorithm we give
  evidence for the presence of strongly anomalous scaling for periodic
  boundary conditions which is much worse than anticipated before.  We
  propose and study alternative boundary conditions for the compass
  model which do not make use of extended configuration spaces and
  show that they completely remove the problem with finite-size
  scaling. In the last part, we apply these boundary conditions to the
  quantum problem and present a considerably improved estimate for the
  critical temperature which should be of interest for future studies
  on the compass model. Our investigation identifies a strong one-dimensional
  magnetic ordering tendency with a large correlation length as the
  cause of the unusual scaling and moreover allows for a precise
  quantification of the anomalous length scale involved.

\end{abstract}

\maketitle

\section{Introduction}
The quantum compass model \cite{kugel82} has recently seen a
renaissance in condensed-matter physics, which was to a large part
triggered by the observation that it may protect q-bits in a quantum
computing setting \cite{doucot:024505,milman:020503}. This observation
may be of actual practical relevance as the quantum compass model can
be realized by a special connection of Josephson junction arrays, a
concept with which first experimental successes could be reported
\cite{gladchenko-2008}. A concrete realization in terms of real
materials has also been proposed recently \cite{jackeli-2008}.  Apart
from the current interest from the quantum information perspective,
the quantum compass model is relevant as an effective description for
orbital ordering, and was originally proposed in this setting
\cite{kugel82}.  Due to the diverse interest in the model, recent
contributions in the literature have studied many different aspects,
ranging primarily from detailed investigations of the ground-state
properties \cite{doucot:024505,dorier:024448} to a study of the
possible low-temperature phases and phase transitions in both the
classical and quantum cases
\cite{mishra:207201,tanaka:256402,wenzelQCMPRB}. Complementary to
that, recent studies considered modified variants of the compass model
in one-dimensional chains \cite{CM1DBrzezicki,eriksson:224424,CM1DSun}
or in a magnetic field \cite{scarola:CM}. In
Ref.~\onlinecite{wenzelPOM}, two of us have proposed and studied a
two-dimensional (2D) geometric variant of the compass model.  The
model is also known to have relevance for other settings such as $p+ip$
superconductors \cite{PhysRevLett.93.047003,nussinovpip}, the concept
of dimensional reduction \cite{Batista_dimred2005}, and was recently
shown to be isospectral \cite{KaiToricTransverse} to Kitaev's toric
code \cite{kitaev-toriccode} in a field.

The 2D compass model (CM) is defined on a square lattice of  $N= L\times L$ sites as a
(pseudo) spin model by the Hamiltonian
\begin{equation}
\label{eq:hamiltonian0}
\mathcal{H}_\mathrm{CM}=J_x\sum_i  S^x_i S^x_{i+\mathbf{e}_x} + J_y \sum_i S^y_i S^y_{i+\mathbf{e}_y}\,,
\end{equation}
where $S^x_i$ and $S^y_i$ are components of a two-component spin
$\mathbf{S}_i$ at site $i$. The spin can represent both classical and
quantum degrees of freedom. In the latter case $S^x$ and $S^y$ are
represented by the usual Pauli matrices, i.e.,
$\mathbf{S}=(1/2)(\sigma_x,\sigma_y)$. The classical case is analogous
to an ordinary classical XY spin $\mathbf{S}=(S^x,S^y)\in S^1$.  The
interesting feature of Eq.~\eqref{eq:hamiltonian0} is its anisotropy in
spin \emph{and} lattice space.

For $J_y \neq J_x$, the ground state of Eq.~\eqref{eq:hamiltonian0}
can be described by (weakly) coupled Ising spin chains oriented in the
$x$ or $y$-direction depending on $|J_x| > |J_y|$ or $|J_y| > |J_x|$,
respectively. The quantum phase transition between these differently
oriented ground states was shown to be of first-order
\cite{CMfirstorder,CMorusfirstorder}. One interesting feature of that
work is that Ref.~\onlinecite{CMorusfirstorder} gives one of the first
nontrivial applications of the recently introduced infinite
pair-entangled tensor product states (iPEPS) \cite{iPEPS} which aim at
providing a new numerical approach to 2D interacting quantum systems.
Following the same line of research, a quantum phase
  transition in a generalized CM has also been investigated recently
\cite{Cincio} using the related multiscale entangled renormalization
ansatz (MERA) \cite{MERA}.

Here, we will focus on the symmetric case $J_x=J_y=-1$ which
allows -- due to a discrete $x \leftrightarrow y$ symmetry in spin and lattice space
\cite{mishra:207201} -- for a thermal phase transition to a
directionally-ordered low-temperature phase without long-range local
magnetic order \cite{mishra:207201, wenzelQCMPRB}. In
Ref.~\onlinecite{wenzelQCMPRB}, two of us have studied this
transition extensively for both the classical and quantum CMs. One of
the main results of this contribution is the confirmation that the CM suffers
from extraordinary finite-size corrections when studied in a simple
canonical ensemble on the torus, contradicting the naive assumption that
generic periodic boundary conditions are optimal. A solution to
alleviate this problem had already been suggested in
Ref.~\onlinecite{mishra:207201}, where the authors proposed the use of
so-called fluctuating boundary conditions (FBC) (sometimes also
referred to as ``annealed boundary
conditions''). These formally place the CM in a larger configuration
space where the partition function is given by
\begin{equation}
\label{eqn:FB}
Z_\mathrm{FBC} = \sum_{\{J_b=\pm 1\}} \int \prod_i \mathrm{d} \mathbf{S}_i \exp(-\beta \mathcal{H}_\mathrm{CM})
\end{equation}
instead of just
\begin{equation}
\label{eqn:PB}
Z_\mathrm{PBC} = \int \prod_i \mathrm{d} \mathbf{S}_i \exp(-\beta \mathcal{H}_\mathrm{CM})\,.
\end{equation}
for the standard canonical ensemble with periodic boundary conditions
(PBC). Here, $\{J_b\}$ denotes the set of boundary bonds on the periodic lattice
which are allowed to fluctuate  between $-1$ and $+1$ individually \footnote{Or one
  arbitrary bond on each row and column of the lattice.}. One assumes
that the $J_b$ degrees of freedom become unimportant in the
thermodynamic limit.  Indeed, it was shown in
Refs.~\onlinecite{mishra:207201} and \onlinecite{wenzelQCMPRB} that FBC lead to very
good finite-size scaling (FSS) properties in the classical case from which we have good
evidence that the directional-ordering (DO) transition in the CM is in
the 2D Ising universality class.

Our good experience with these FBC is unfortunately of no use for the
quantum CM because of the minus-sign problem. Furthermore, simulations
of the quantum CM are quite demanding and one may currently not reach
large lattice sizes (say $L>64$) with reasonable effort. In result,
our current estimate of the critical ordering temperature is not very
precise as it rests on the use of non-optimal boundary conditions on
moderate lattice sizes \cite{wenzelQCMPRB}. Yet, given the large
interest in the model we find it valuable to try to improve the
accuracy of the critical temperature. A better knowledge of such
quantities is necessary in order to tackle more advanced features such
as the influence of disorder, etc. \cite{tanaka:256402}. Apart from
the motivation to improve the available critical data, there are
further unsatisfactory points or open problems regarding the previous
results \cite{mishra:207201, wenzelQCMPRB}. These especially concern
the \emph{ad hoc} use of FBC to get precise results at the price of
introducing extra degrees of freedom to the model. Why do these
boundary conditions work so well and why do we observe a complete
failure of the critical Binder parameter on periodic lattices?  In
this work we (re)address those questions and present improved results
on critical properties of the classical and quantum CM that we
obtained with a combination of algorithmic advances and by employing
so-called screw-periodic boundary conditions.

The outline of the rest of the paper is as follows. In Sec.~\ref{sec:MC} we start with a revision of
our Monte Carlo (MC) approach and present an improved MC algorithm
building on the Wolff cluster method. The latter will make possible a
much more detailed comparison of FSS properties on periodic vs.
fluctuating boundary conditions in Sec.~\ref{sec:CM}, going
considerably beyond Ref.~\onlinecite{wenzelQCMPRB}. We will show that
periodic boundary conditions behave even worse than previously
anticipated. A solution to this problem is thereafter suggested in
form of screw-boundary conditions.  These will allow to recover very
good scaling properties without making use of an extended
configuration space (in form of fluctuating boundary conditions).
Moreover they can be readily employed in quantum Monte Carlo (QMC) simulations which
is the topic of Sec.~\ref{sec:QCM}, where improved critical data for
the quantum CM are presented. We end with a summary and conclusions in
Sec.~\ref{sec:concl}.

\section{\label{sec:MC}Observables and MC approach}
In this section, we present the standard approach to simulate the
classical CM and describe in detail how we can improve the algorithm
by making use of ideas from well-known cluster MC updates. A short
discussion of the QMC approach for the quantum version of
Eq.~\eqref{eq:hamiltonian0} is postponed to Sec.~\ref{sec:QCM}.

\subsection{\label{sec:MC:A}Revision of classical MC approach and relevant observables}

In the classical case, we have investigated the ensembles specified by
Eqs.~\eqref{eqn:FB} and \eqref{eqn:PB} using the Metropolis algorithm
combined with the parallel tempering (PT) scheme
\cite{geyerPT,hukushimaPT,BittnerPT} parallelizing
  simulations at different temperatures. Technical details of this
approach are described in Ref.~\onlinecite{wenzelQCMPRB}. During a MC
simulation, we measure an order-parameter known to describe
directional-ordering
\begin{equation}
D=\frac{1}{N} | E_x- E_y |\,,
\end{equation}
with $E_x=\sum_{i} S_i^x S^x_{i+\mathbf{e}_x}$ and $E_y=\sum_{i} S_i^y
S^y_{i+\mathbf{e}_y}$. We concentrate here on its susceptibility
\begin{equation}
\chi=N\left(\langle D^2 \rangle  - \langle D \rangle ^2 \right)\,,
\end{equation}
which diverges at the phase transition temperature $T_\mathrm{c}$. On
finite systems the divergence in $\chi$ is smoothened into a finite
maximum $\chi_{\max}(L)$ at some pseudocritical temperature
$T_{\max}(L)$.  Finite-size scaling predicts the following two
fundamental scaling relations (see, e.g.,
Refs.~\cite{vicarireview,LandauMCBook,JankeMCGreifswald})
\begin{align}
\label{eqn:FSSchi}
\chi_{\max}(L) & \sim L^{\gamma/\nu}\,,\\
\label{eqn:FSST}
T_{\max}(L) & =T_\mathrm{c}+aL^{-1/\nu}\,,
\end{align}
which are the primary means used in this paper to obtain the critical
exponents $\nu,\gamma$ and the critical temperature $T_\mathrm{c}$, and to discuss anomalous
scaling. From MC simulations at discrete temperatures in the vicinity
of the phase transition, we obtain $\chi_{\max}(L)$ and $T_{\max}(L)$
by making use of standard reweighting techniques
\cite{ferrenberg:multi} and optimization algorithms. Error estimates
for these quantities are obtained by ``jackknifing'' this procedure
\cite{efron,JankeMCGreifswald}.

\subsection{\label{sec:MC:Wolff}A Wolff-like cluster update}

Up to now the application of the PT technique has proven to be quite
efficient, enabling a study of the classical (and quantum) CM on moderately large lattice
sizes \cite{wenzelQCMPRB}. However, for linear system sizes of about $L\gsim 100$, we observe
that the method runs into problems as the equilibration times in the
MC simulation become visibly very long. In order to go
efficiently beyond such lattice sizes, a further improved method is
therefore called for.

Indeed, as we shall propose here, a rather straightforward improvement is possible with a
special one-dimensional Wolff-cluster update \cite{WolffCluster}. To
see this reconsider the ordinary Wolff-construction for $O(N)$ spin
models with Hamiltonian $\mathcal{H}_{O(N)}=J\sum_{\langle ij \rangle}
\mathbf{S}_i \mathbf{S}_j$. Following Ref.~\onlinecite{WolffCluster},
the operation \begin{equation}R^\mathbf{r}(\mathbf{S}_i) =
  \mathbf{S}_i -
  2(\mathbf{S}_i\cdot\mathbf{r})\mathbf{r}\end{equation} denotes a
reflection of the $[O(N)]$ spin $\mathbf{S}_i$ along a hyperplane
defined by the vector $\mathbf{r}$. Given a random $\mathbf{r}$, Wolff
clusters are constructed using the bond activation probability
\begin{equation}
\label{eqn:wolffacc}
P_{ij} = 1 - \exp\left( \min [0, -J\beta [ \mathbf{S}_i \mathbf{S}_j- \mathbf{S}_i R^\mathbf{r}(\mathbf{S}_j)] \right)
\end{equation}
for bonds $\langle i j \rangle$. The spins in each cluster are then
flipped by applying $\mathbf{S}_i \to R^\mathbf{r}(\mathbf{S}_i)$
which implements the (non-local) MC move.
The principle of detailed balance is satisfied by requiring
the invariance
\begin{equation}
\label{eqn:condition}
H (R^\mathbf{r}(\mathbf{S}_i), R^\mathbf{r}(\mathbf{S}_j)) = H( \mathbf{S}_i, \mathbf{S}_j) \equiv H_{ij}
\end{equation}
of the bond energy $H_{ij}$ ($\mathcal{H}=\sum_{\langle ij \rangle} H_{ij}$) under
reflection of the spins for each bond $\langle ij \rangle$ of the
lattice. While this is true for $\mathcal{H}_{O(N)}$, it
is clearly not true for the CM in general. However, we know that for
the CM, the following special reflection operations $R^{\mathbf{e}_x}$
and $R^{\mathbf{e}_y}$ with
\begin{align}
R^{\mathbf{e}_y}(S^x_i , S^y_i) &= (-S^x_i, S^y_i),\\
R^{\mathbf{e}_x}(S^x_i , S^y_i) &= (S^x_i, -S^y_i)
\end{align}
are symmetries on a subset of all bonds $\langle i,j \rangle$ \cite{doucot:024505,dorier:024448}, namely that
\begin{align}
\label{eqn:clustersymm}
H(R^{\mathbf{e}_y}(\mathbf{S}_i), R^{\mathbf{e}_y}(\mathbf{S}_{i+\mathbf{e}_x})) &= H(\mathbf{S}_i, \mathbf{S}_{i+\mathbf{e}_x}),\\
H(R^{\mathbf{e}_x}(\mathbf{S}_i), R^{\mathbf{e}_x}(\mathbf{S}_{i+\mathbf{e}_y})) &= H(\mathbf{S}_i, \mathbf{S}_{i+\mathbf{e}_y}).
\end{align}
Thus, we may employ $R^{\mathbf{e}_x}$ and $R^{\mathbf{e}_y}$ to
construct one-dimensional clusters of spins along the $x$- or
$y$-direction. Employing the form of the CM Hamiltonian \eqref{eq:hamiltonian0} and the
general relation \eqref{eqn:wolffacc} we obtain the following
bond-activation probabilities
\begin{align}
P_{i i+\mathbf{e}_x} &= 1-\exp(-2 J_x \beta S^x_i S^x_{i+\mathbf{e}_x}),\\
P_{i i+\mathbf{e}_y} &= 1-\exp(-2 J_y \beta S^y_i S^y_{i+\mathbf{e}_y}) 
\end{align}
for cluster growth along the $x$- and $y$-directions, respectively.
Note that the cluster construction is really strictly one-dimensional,
i.e., when we build $x$-clusters we do not attempt to add $y$-bonds to
the cluster which would break condition \eqref{eqn:clustersymm}.
Cluster construction starts as usual by picking a random start site
from which cluster growth proceeds.

\begin{figure}
\centering
\includegraphics[width=0.45\textwidth]{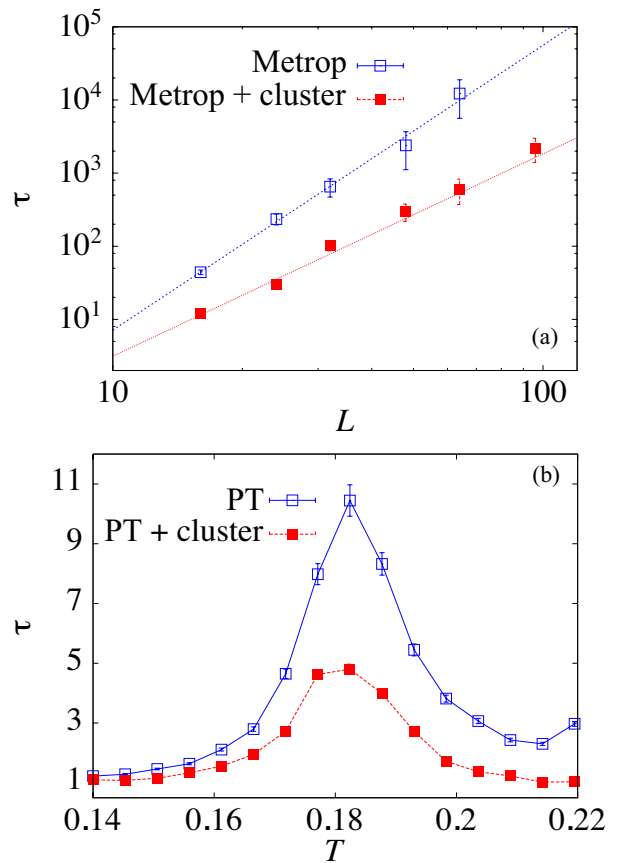}
\caption{\label{fig:Wolffcmp}(Color online) (a) Comparison of the autocorrelation
  time $\tau$ at the critical point using periodic boundary
  conditions.  Comparison of $\tau(L)$ for the pure Metropolis and the 
  combined Metropolis + cluster update. An overall improvement for $\tau$ as
  well as better scaling is evident for the cluster variant. (b)
  Comparison of $\tau$ for $L=36$ as obtained from the pure parallel
  tempering approach with the improved parallel tempering variant.}
\end{figure}

An obvious difference to the original Wollf algorithm is the discrete
set of possible spin reflections. Thus the cluster update alone does
not satisfy ergodicity. This is not a problem as long as ordinary
Metropolis (as well as PT updates) are performed in addition.  In each
MC sweep we perform on average $L$ cluster updates in both $x$ and $y$
directions as well as $N$ local Metropolis updates.  We have verified
by detailed comparison to existing data that the new algorithm works
correctly. Let us proceed directly to an evaluation of the new update. In order to
examine its performance we ran several tests on lattice sizes $L =
16,24,32,48,64,(96\text{ in case of the cluster update})$ in the ensemble $Z_\mathrm{PBC}$ at the pseudocritical
temperatures $T_{\max}(L)$ (known from our previous study). In the first
test we switched off the PT update and compared the autocorrelation
time $\tau$ of the energy time-series which should scale at the
critical point like $\tau \sim L^z$. Figure \ref{fig:Wolffcmp}(a)
compares the scaling of $\tau$ with and without the above cluster
updates.
Clearly, we find that the cluster algorithm behaves much better. Apart
from the expected absolute reduction of $\tau$ we observe a decrease
of $z$ from $z \approx 3.5$ to $z \approx 2.5$ which is apparent from
the different slopes in the log-log plot.  Next, we also compared the
autocorrelation time $\tau$ in simulations employing the PT algorithm.
Without performing a detailed scaling analysis, it is evident from
Fig.~\ref{fig:Wolffcmp}(b) that the cluster update further improves the PT
algorithm.

The methodological improvement presented here allows to study much
larger system sizes than before. In the course of this study, we have
performed simulations up to $L=512$. 

\section{\label{sec:CM}Classical Compass Model: Results}

In this section we are going to employ the algorithmic advances to
restudy critical properties of the classical CM. Special focus is given to
a more detailed comparison of ensembles $Z_\mathrm{FBC}$ and
$Z_\mathrm{PBC}$. Based on this comparison we will thereafter propose
the use of alternative boundary conditions and study their effects
on FSS properties.

\subsection{\label{sec:CM:A}Revisiting periodic and fluctuating boundary conditions}

Previous investigations of the DO transition have
clearly shown that the use of ensemble $Z_\mathrm{FBC}$ is favorable
over $Z_\mathrm{PBC}$ in terms of FSS properties
\cite{mishra:207201,wenzelQCMPRB}, where the most severe ``failure'' of
$Z_\mathrm{PBC}$ establishes itself in an unconventional behavior of
the Binder parameter. Despite these observed defects it was argued
\cite{wenzelQCMPRB} that one may still employ PBC to extract the
critical properties given the system sizes $L$ are large enough. This
argument was supported from extrapolations of pseudocritical
temperatures $T_{\max}(L)$ which gave consistent results for both $Z_\mathrm{FBC}$ and
$Z_\mathrm{PBC}$ of about $T_\mathrm{c}=0.1464(2)$.

With the newly available cluster procedure, we will investigate
this issue further to make more quantitative statements about how
ensembles $Z_\mathrm{FBC}$ and $Z_\mathrm{PBC}$ converge towards another
asymptotically.  We have thus simulated the CM for system sizes
between $L=12$ and $L=512$, pushing $L$ a factor of $5-10$ times
larger than before. In comparison to Ref.~\onlinecite{wenzelQCMPRB}, we
have added system sizes $L=96,128,164,256,512$.

The observables described in Sec.~\ref{sec:MC:A} were estimated using
about $10^5$ samples. We have taken measurements only every $m$ MC
sweep such that the final autocorrelation time was small, $\tau < 10$ 
($m$ was in the range of $4-100$).
\begin{figure}[h!]
\centering
\begin{minipage}{0.49\textwidth}
\includegraphics[width=\textwidth]{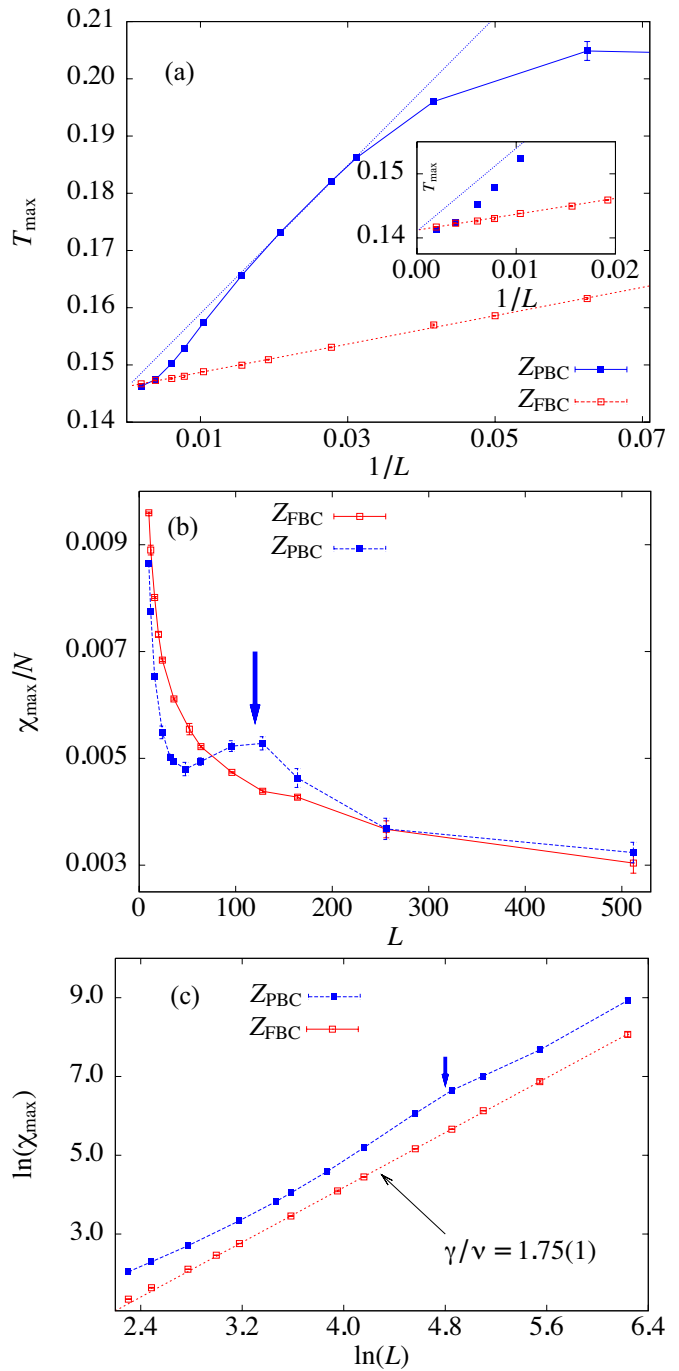}
\end{minipage}
\caption{\label{fig:results1}(Color online) Improved finite-size data
  for the classical compass model with periodic (PBC) and fluctuating
  boundary conditions (FBC).  (a) Extrapolation of the pseudocritical
  temperatures $T_{\max}(L)$. Data for FBC follow a perfect straight
  line. The periodic case shows a clear bend for system sizes $L>96$
  not anticipated before based on the straight line extrapolation in
  Ref.~\cite{wenzelQCMPRB}. (b) The susceptibility maxima divided by
  $N$ behave non-monotonously and indicate a resonance phenomenon at
  about $L\approx 120$ (indicated by the arrow). (c) FSS of
  $\chi_{\max}$ in a log-log plot. For FBC a power law is evident with
  $\gamma/\nu=1.75(1)$. For PBC a power-law extrapolation is not
  justified and different scaling regimes are apparent.}
\end{figure}
For a presentation of the typical temperature dependence of the order
parameter and susceptibility  we refer the reader to the previous
work of Ref.~\onlinecite{wenzelQCMPRB}. Here we just present the
pertinent data obtained for the pseudocritical temperature
$T_{\max}(L)$ and $\chi_{\max}(L)$. Figure \ref{fig:results1}
summarizes the FSS analysis for the two different ensembles
considered. The partly surprising results of this comparison can be
summarized as follows.

First, we observe that the FSS behavior for $Z_\mathrm{FBC}$
is fully consistent with earlier results, i.e., data obtained on
larger lattices agree with the extrapolations from smaller
lattice sizes. This further confirms the claim of 2D Ising universality 
beyond any reasonable doubt. Indeed, fits to Eq.~\eqref{eqn:FSST}
yield our estimate of the critical temperature and critical exponent
as
\begin{equation}
\label{eqn:Tccl}
T_\mathrm{c}=0.14621(2),\quad \nu=1.00(1)
\end{equation}
(with $\chi^2/\mathrm{d.o.f}=1.4$) which represents an improvement of
roughly one order of magnitude over the previous estimate. Together
with the critical exponent $\gamma=1.75(1)$ obtained from analyzing
the scaling of $\chi_{\max}$, this is in perfect agreement with the exactly
known critical exponents for the 2D Ising model.

Second --- and this is the surprising result --- the scaling for
$Z_\mathrm{PBC}$ reveals a more complicated or stronger \emph{anomalous
  scaling} than previously thought. This is especially apparent in
Fig.~\ref{fig:results1}(a) where pseudocritical temperatures for $L>96$
clearly deviate systematically from the previous extrapolation [upper (blue) straight line] of Ref.~\onlinecite{wenzelQCMPRB} based on the assumption of 2D
Ising scaling for $L > 30$. Note that the upper (blue) line was also
justified because it matched exactly with the result from
$Z_\mathrm{FBC}$. However, instead of a clear power law scaling in
$1/L$ we observe a ``double bend'' in the FSS curve which seems to
collapse onto the curve from $Z_\mathrm{FBC}$ for very large system
sizes $L \geq 256$ (inset of Fig.~\ref{fig:results1}(a)). 

The same anomalous scaling behavior shows up in the quantity
$\chi_{\max}(L)/N$ of Fig.~\ref{fig:results1}(b) which shows a strong
non-monotonic behavior at a length scale of about $L\approx 120$.
Thus, any attempt to extract the critical exponent $\gamma$ from a
log-log plot as in Fig.~\ref{fig:results1}(c) is doomed to fail on length scales below
$L\approx 256$. The observed non-monotonic behavior also shows up in the Binder
parameter but we postpone a discussion on that to the next subsection.

As a matter of fact, it is thus totally unreliable to obtain critical
exponents and the critical ordering temperature $T_\mathrm{c}$ from
simple extrapolations in the ensemble $Z_\mathrm{PBC}$ on periodic
lattices (at least for $L\lsim 256$). The previous seemingly correct
extrapolation was a matter of coincidence.  Turning this observation
around, one might even be tempted to argue for non-Ising behavior in
the CM if one did not have access to the largest lattice sizes studied
here. This situation is most unsatisfying and calls for a deeper
investigation and a workaround. We will attempt precisely this in the next
subsection.

\subsection{\label{sec:CM:B}Screw-periodic boundary conditions}

The main message of the discussion so far is that PBC show a more
complex scaling behavior than previously thought with the appearance
of a clear resonance effect and at least two different scaling
regimes. This disqualifies the use of PBC to extract critical
properties. The FBC ensemble on the other hand also has --- despite its
intriguing performance --- a couple of drawbacks. The most important
of all is that we may not easily use it in QMC because fluctuating
couplings induce a minus-sign problem. Second, one may wonder whether
it is safe to use them in the first place as a trustable FSS theory
is not available and one is actually simulating a different model.

Here, we would like to ask whether it is nevertheless possible to deal
with the described problem using only slightly modified boundary
conditions without going to a higher dimensional configuration space.

\begin{figure}
\begin{minipage}{0.23\textwidth}
\includegraphics[width=0.95\textwidth]{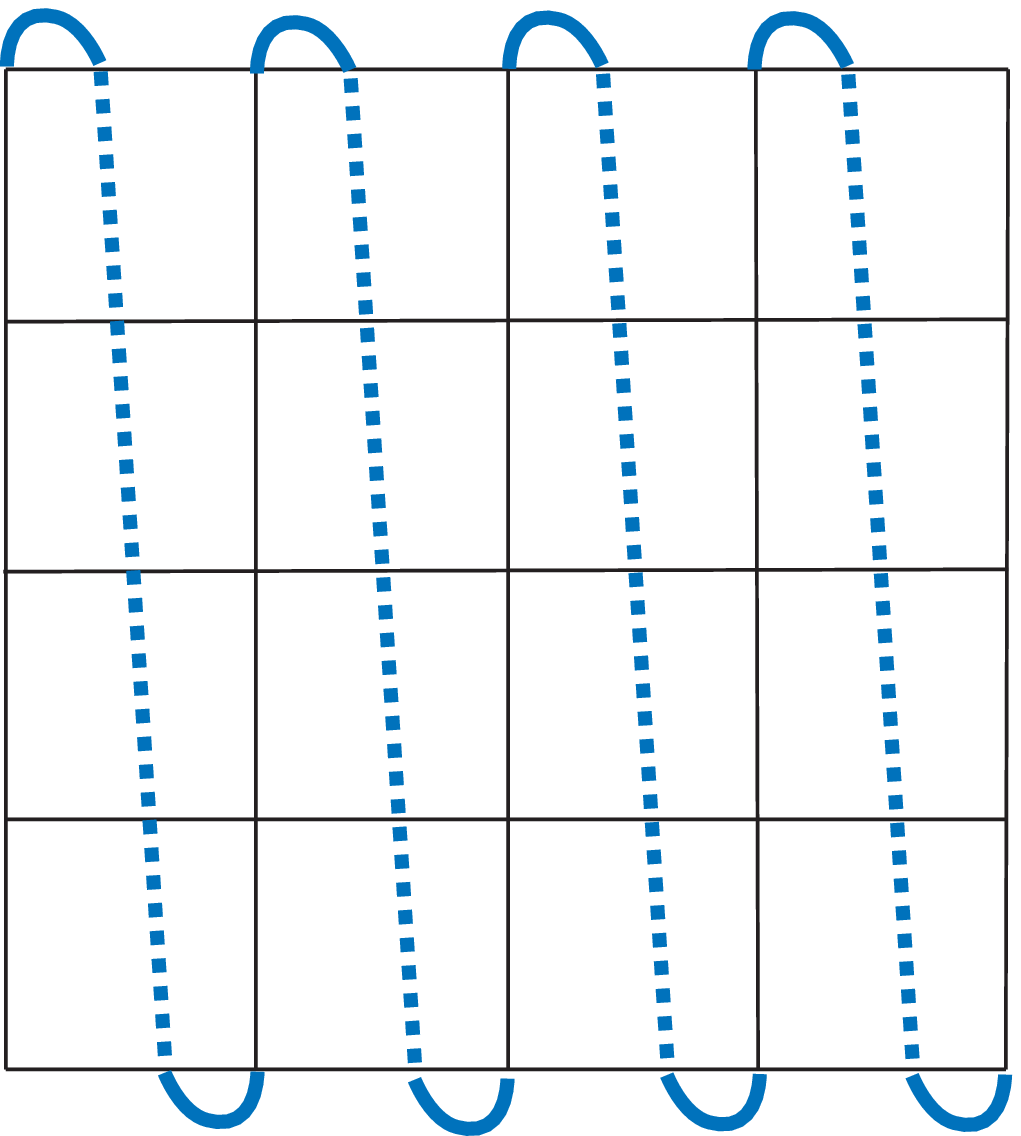}
\end{minipage}
\begin{minipage}{0.23\textwidth}
\includegraphics[width=0.95\textwidth]{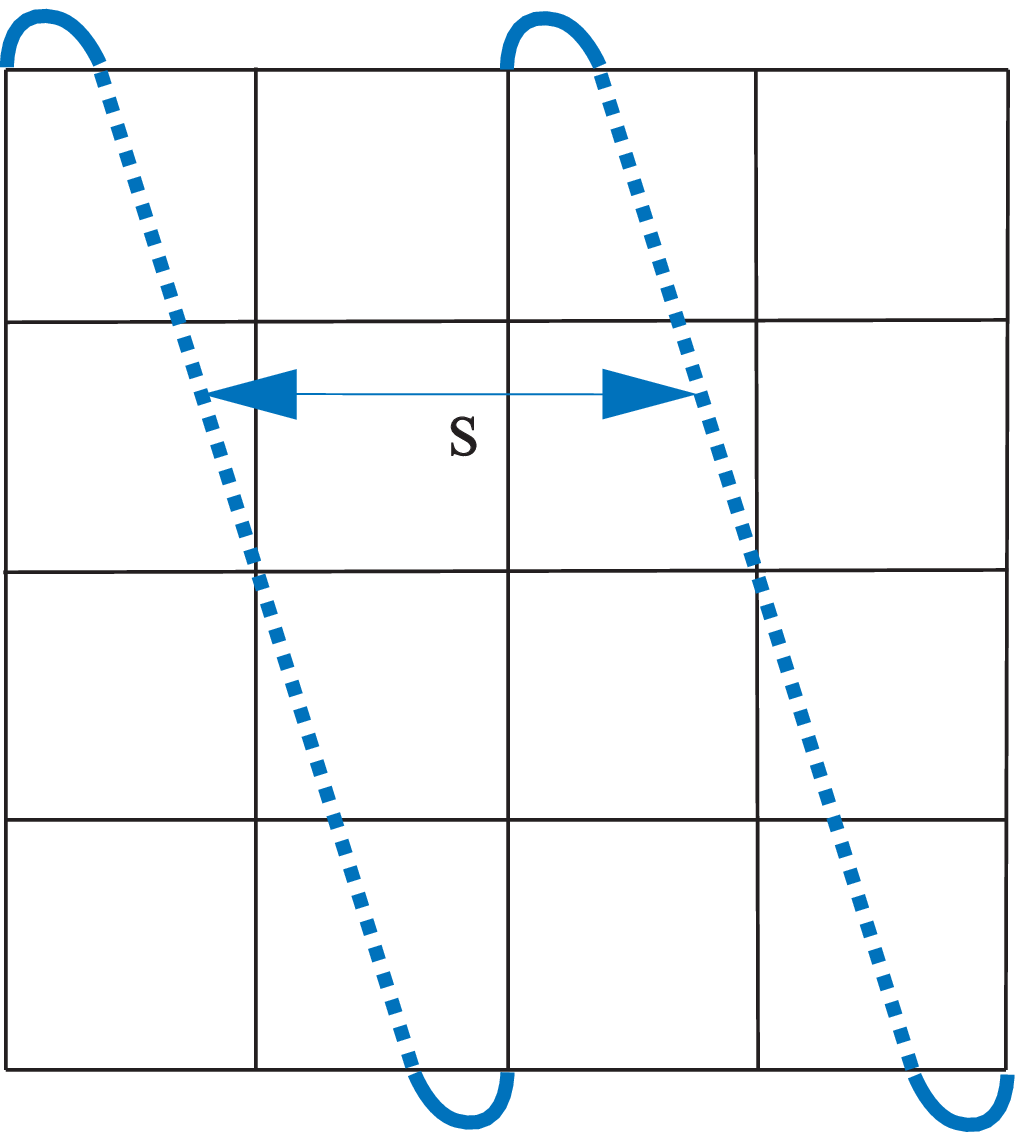}
\end{minipage}
\caption{\label{fig:screw}(Color online) Illustration of
  screw-periodic boundary conditions along the $y$ axis as defined in
  Eq.~\eqref{eqn:screwbc}. Two examples with $S=1$ and $S=2$ are
  shown. In our simulations the same procedure is applied to the
  $x$-direction.}
\end{figure}

It is quite obvious that the torus geometry hides or shields the true
physics going on.  One possibility to unveil the true properties of
the CM in the thermodynamic limit is to introduce systematic
deformations to the torus. In this way we can at least see how the
problem is alleviated (or made worse). Among all such deformations one
may consider a M\"obius strip or so-called screw-periodic boundary
conditions. Such deformations are very easy to implement on the
computer and cost no extra updates. We decide to study screw-periodic
boundary conditions (SBC) which are defined by
\begin{align}
\label{eqn:screwbc}
(x,y+1)&=\begin{cases} (x,y+1) & \text{if}\quad y<L-1 \\ ([x+S]\!\mod L,0) & \text{if}\quad y=L-1,\end{cases}\\
\nonumber
(x+1,y)&=\begin{cases} (x+1,y) & \text{if}\quad x<L-1 \\ (0,[y+S]\!\mod L) & \text{if}\quad x=L-1,\end{cases}  
\end{align}
where, e.g., $(x,y+1)$ denotes the nearest-neighbor of site $i=(x,y)$
in $y$-direction. The parameter $S$ is a parameter that determines how
much we deform the clean torus case. Figure~\ref{fig:screw}
illustrates this concept for two cases $S=1$ and $S=2$. The cases
$S=0$ and $S=L$ are obviously identical to the usual PBC.
For a given lattice size, $S$ may take only certain values in order to
satisfy the overall periodicity constraint. The possible $S$-values
are given by the set of all (distinct) divisors of $L$. SBC are
discussed in various forms in the literature, sometimes called helical
boundary conditions or shift boundary conditions. Mostly they have
been employed because they have some advantages regarding
implementation issues \cite{LandauMCBook,BarkemaMCBook} or to
complement FSS analysis \cite{Ising_screwbc} as they approach the
thermodynamic limit with (slightly) different pseudocritical
temperatures. A further useful application is the controlled
formation of \emph{tilted} domain walls in the Ising model
\cite{BittnerscrewBC}. Note that each site still has exactly four
neighbors which distinguishes SBC from open boundary conditions.

SBC allow one to put the lattice points into representation classes which
we will call loops. A loop is the set of all points $i$ that the
screw/helix passes until it closes itself. The length of a loop is
called $L_l$ and is the number of points it contains. Each point $i$
is obviously member in exactly two loops, one for the $x$- and one for
the $y$-direction. Given a lattice size $L$ and compatible screw
parameter $S$, each loop has length $L_l = L^2/S$ and for $S=1$ all
points belong to the same loop. 
The notation introduced here becomes
relevant when discussing the symmetry properties of the CM and its
ground-state degeneracies.

A simple check confirms that the usual one-dimensional spin flip
operators $P_l=\prod_x \sigma^y_{(x,l)}$ and $Q_m=\prod_y \sigma^x_{(m,y)}$
\cite{doucot:024505,dorier:024448} (which are related to operations
$R^{\mathbf{e}_x}$ and $R^{\mathbf{e}_y}$ in Sec.~\ref{sec:MC:Wolff})
are no longer symmetries of the (quantum) Hamiltonian
\eqref{eq:hamiltonian0} if $S\neq 0$. However they can be generalized to the SBC case with the following operators
\begin{align} 
P_l&= \prod_{j \in l} \sigma^y_j,\\ 
Q_m&= \prod_{i \in m} \sigma^x_i,
\end{align}
where $l$ and $m$ now refer to a loop along the $x$ or $y$-direction.
As we can control the number of independent loops via the parameter
$S$, we can control the number of such symmetry operators and thus the
degeneracy of the ground-state. Indeed it is possible to change the
ground-state degeneracy from exponential growth $2^{L+1}$ ($S=0$) to a
constant $2$ ($S=1$), an observation which may have interesting
physical consequences. The thermal DO transition
studied here should not be affected by this as the relevant global
$Z_2$ symmetry is not changed.
\begin{figure}
\centering
\begin{minipage}{0.49\textwidth}
\includegraphics[width=\textwidth]{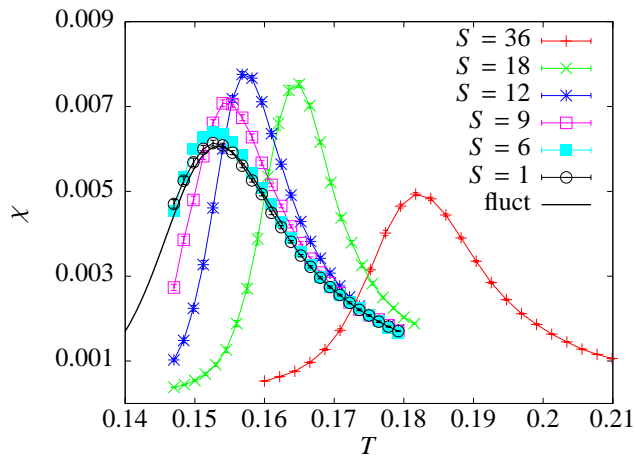}
\end{minipage}
\caption{\label{fig:screw_suscmp}(Color online) Dependency of the
  susceptibility $\chi(T)$ for $L=36$ on the choice of the screw
  parameter $S$. The case $S=36$ is equivalent to periodic boundary
  conditions. With decreasing $S$ (or increasing the boundary loop
  length) a clear shift in the peak is observed with an apparent
  resonance at $S \approx 12$. For $S=1$ the susceptibility is
  (nearly) identical to the susceptibility obtained in the fluctuating
  bond ensemble $Z_\mathrm{FBC}$ (continuous line without symbols).}
\end{figure}

Let us proceed to study the effect of SBC in actual MC simulations.
To this end, we choose a system size $L=36$ which allows us to study
quite a large number of screw parameters $S=0,1,2,3,4,6,9,12,18$. In
each case we measured the order parameter $D$ and its susceptibility
$\chi$ for a couple of temperatures close to the phase transition.
Figure~\ref{fig:screw_suscmp} depicts the drastic effect of SBC on the
susceptibility $\chi$. Starting from the periodic case $S=0$ (or $S=36$) we
observe that $\chi$ moves massively towards the curve from
$Z_\mathrm{FBC}$ for decreasing $S>0$ or increasing $L_l$. The case
$S=1$ gives an almost identical result to that obtained with
fluctuating couplings in the $Z_\mathrm{FBC}$ ensemble. Second it is
apparent that there is a resonance at some length scale determined by
$S\approx 12$ at which the fluctuations in the system are strongest.

\begin{figure}
\centering
\begin{minipage}{0.49\textwidth}
\includegraphics[width=\textwidth]{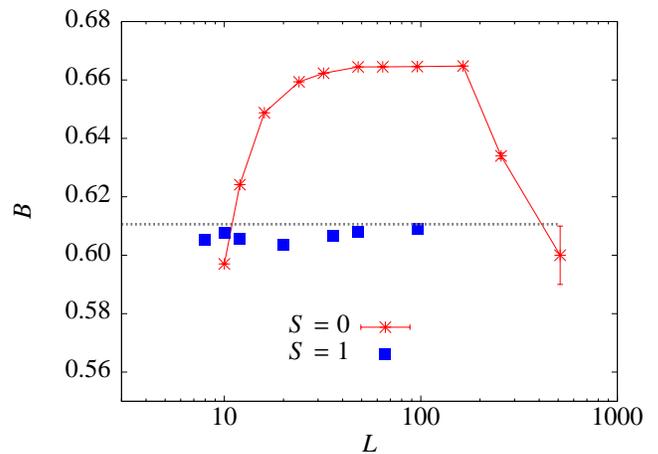}
\end{minipage}
\caption{\label{fig:Bcomparison}(Color online) Finite-size behavior of
  the Binder parameter $B$ at the critical point $T=T_{\mathrm{c}}$
  for periodic ($S=0$) and screw boundary conditions with $S=1$. While
  there is a strong anomaly for $S=0$, the case $S=1$ almost
  completely removes the defect and is consistent with a scale
  independent value of $B$ at the critical point. Moreover, it is
  consistent with the usual 2D Ising value
  \cite{Kamieniarz,SalasBinderIsing} indicated by the horizontal
  line. Note that the symbol size for $S=1$ is bigger than the error bar.}
\end{figure}

The above picture thus suggests that $S=1$ resolves the FSS problems
observed in the CM for the susceptibility almost completely.
Furthermore it gives a hint at the order of the disturbing length
scale ($L_l \approx 36^2/12 \approx 110$) which is present and which
prohibits the extraction of correct critical data.  Any solution that
restores good FSS properties should also repair the behavior of the
Binder parameter
\begin{equation}
B=1-\frac{1}{3}\frac{\langle D^4 \rangle}{\langle D^2 \rangle^2},
\end{equation}
whose normally used power is due to a scale invariance at the critical
point with only leading order corrections. Thus, if SBC really solve
the problem they should also remove the very unconventional behavior
in the Binder parameter which was observed with PBC in
Ref.~\onlinecite{wenzelQCMPRB}. Figure \ref{fig:Bcomparison} shows a
comparison of the finite-size behavior of $B$ for the cases $S=0$
(periodic) and $S=1$ performed close to the critical point given in
Eq.~\eqref{eqn:Tccl}. The periodic case shows the expected
non-monotonous behavior (with a possible restoration for $L>256$).
The $S=1$ screw restores the expected scaling behavior -- up to a
small bump for $L<12$ -- completely, i.e., it is almost a constant for
various system sizes and agrees rather well with the known value of
$B\approx 0.61$ \cite{Kamieniarz, SalasBinderIsing} for the 2D Ising
model (constant line in Fig.~\ref{fig:Bcomparison}). Note, however, that
the agreement is not expected to be perfect as boundary conditions can
have (a small) influence on the (only weakly universal) critical value of $B$
\cite{SelkeBinderBC}. For an analysis of the Binder parameter for FBC we refer the reader to Ref.~\cite{wenzelQCMPRB}.

This brings us into the position to claim that SBC are a very
efficient tool to study critical properties of the CM. Before we apply
these to the quantum CM let us try to shed some light onto the origin
of anomalous scaling (with PBC).

\subsection{Origin of anomalous scaling: one-dimensional spin order}

It is evident from the MC analysis in Secs.~\ref{sec:CM:A} and
\ref{sec:CM:B} that there is a second length scale in the CM which
influences fluctuations and which can be overcome by SBC. Let us now
turn to a discussion of possible reasons for this as a more
fundamental understanding of this phenomenon is clearly desirable.

We know that the low-$T$ (directionally-ordered phase) of the CM is
essentially one-dimensional where the spins along each row  or column
are essentially decoupled. These spins thus form a 1D spin chain.
Using this picture, a plausible explanation for the failure of FSS was
actually suggested in Ref.~\onlinecite{mishra:207201} where it is
argued that the \emph{magnetic} spin-spin correlation length
$\xi_{1D}$ along each chain exceeds the linear system size $L$ at
low temperatures.
If this were the case, all spins would align themselves along each
chain although a directionally ordered state can survive even with
domain walls in spin space. Such total magnetic ordering tendency could
influence the fluctuations of the true order parameter, making it more
robust against thermal fluctuations and spoiling its FSS properties.

To test this hypothesis let us write down an order parameter for such
one-dimensional magnetic ordering tendency $M_{1D}$ as
\begin{align}
M^{x(y)}_{1D} &= \frac{1}{N_L} \sum_{\mathrm{x(y)-loops}, l} \frac{1}{L_l} \left| \sum_{i\in l} S_i^{x(y)} \right|,\\
M_{1D} &=\frac{1}{2}(M^x_{1D} + M^y_{1D}). 
\end{align}
Here $N_L$ denotes the number of boundary loops (as introduced above)
and $L_l$ the length (number of sites) of loop $l$, i.e., we
already take care for the general screw-periodic case. The quantity
$M_{1D}$ probes whether all spins along each chain (or loop) like to
align themselves.

To test whether such possible (long-range) ordering of the spins
exists on top of directional order, we perform a couple of MC runs
at the critical temperature $T_\mathrm{c}$ obtained in Eq.~\eqref{eqn:Tccl}
for lattice sizes $L=8,12,20,36,48$. In each case we simulate all
possible screw parameters $S$.
\begin{figure}
\centering
\begin{minipage}{0.49\textwidth}
\includegraphics[width=\textwidth]{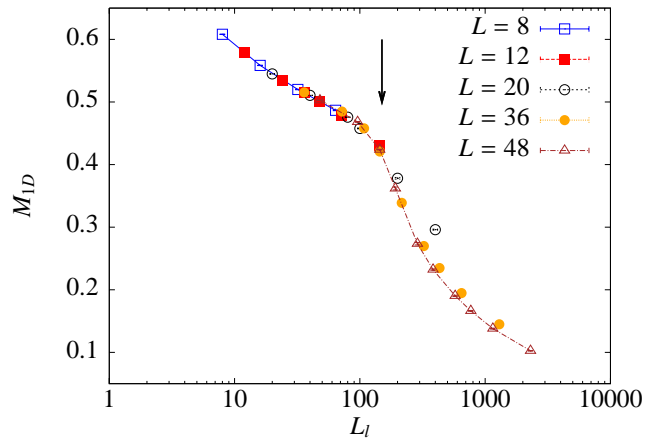}
\end{minipage}
\caption{\label{fig:MAG1D}(Color online) Expectation values for
  one-dimensional magnetization for several lattice sizes and choices
  of the screw displacement $S$ at the critical temperature
  $T_\mathrm{c}$.  The $x$-axis is the loop length $L_l=L^2/S$. Data
  from different lattice sizes collapse onto one curve (for $L_l \lsim
  100$). A clear crossover from a region with finite magnetization to
  a disordered spin state is observed on a length scale $L_l \approx
  100$ (indicated by the arrow). This length scale corresponds to
  those where anomalies are seen in the FSS analysis.}
\end{figure}
In Fig.~\ref{fig:MAG1D} we plot the expectation values of $M_{1D}$
\emph{vs} the screw loop length $L_l=L^2/S$. Remarkably, the data from
different system sizes collapse onto the same curve for $L_l \lsim
100$ where a finite expectation value for $M_{1D}$ is evident. This
magnetic ordering tendency does not persist in the thermodynamic limit as for
$L_l \gsim 100$ it suddenly approaches $0$. We conclude that there is
a strong tendency for the spins to align themselves which is enforced
by PBC. Application of SBC can overcome this problem because it
exceeds the typical length scale along each loop. The same is true for
FBC by artificially introducing kinks in the spin configurations,
which is the basic reason why they do not show anomalies such as those in
Fig.~2(b).

The length scale $L_\mathrm{c}$ at which the sudden decrease in $M_{1D}$ appears
coincides precisely with the non-monoticities observed in the scaling
of $T_{\max}$ and $\chi_{\max}$ of Fig.~\ref{fig:results1}. The
resonance effect in Fig.~\ref{fig:screw_suscmp} can be explained
because at $L_l = L_\mathrm{c}$ we have strong fluctuations in
$M_{1D}$ in addition to the normal fluctuations in the directional-order
parameter $D$.

These results essentially confirm the picture of
Ref.~\cite{mishra:207201} and quantify precisely the length scale
involved. The quite large magnetic correlation length can be
understood by recalling the exponential divergence of the magnetic
correlation length at low temperatures in the 1D Ising model.

\section{\label{sec:QCM}Results for the Quantum Case}

We have now developed everything to proceed to the main objective of
this paper which is to improve the estimate of the critical ordering
temperature $T_\mathrm{c}$ for the DO transition in
the presence of quantum fluctuations. Due to the results of
Sec.~\ref{sec:CM:A}, it is probable that the previous result
$T_\mathrm{c}=0.055(1)$ in Ref.~\cite{wenzelQCMPRB} is slightly off the true
critical temperature due to the presence of the second length scale.

\begin{figure}
\begin{minipage}{0.49\textwidth}
\includegraphics[width=\textwidth]{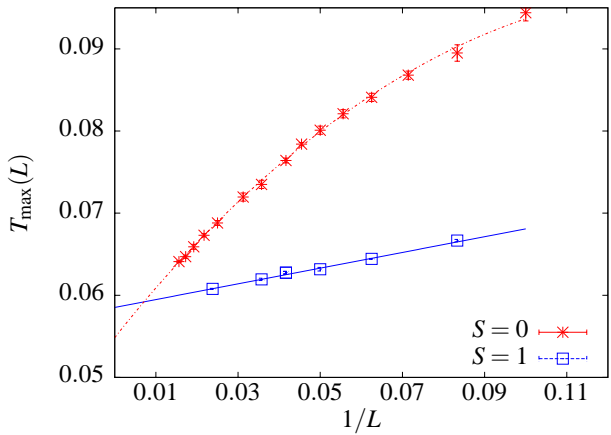}
\end{minipage}
\caption{\label{fig:ffs}(Color online) FSS plot of the
  pseudocritical temperatures for the quantum CM from the susceptibility comparing
  periodic ($S=0$) and screw-periodic boundary conditions ($S=1$). The latter
  clearly lead to a considerable improvement. The line is a fit using
  a power law correction in $1/L$. The dashed extrapolation of data
  from periodic boundary conditions \cite{wenzelQCMPRB} underestimates
  the critical temperature as expected from the discussion in
  Sec.~\ref{sec:CM:A}.}
\end{figure}

\begin{figure}
\begin{minipage}{0.49\textwidth}
\includegraphics[width=\textwidth]{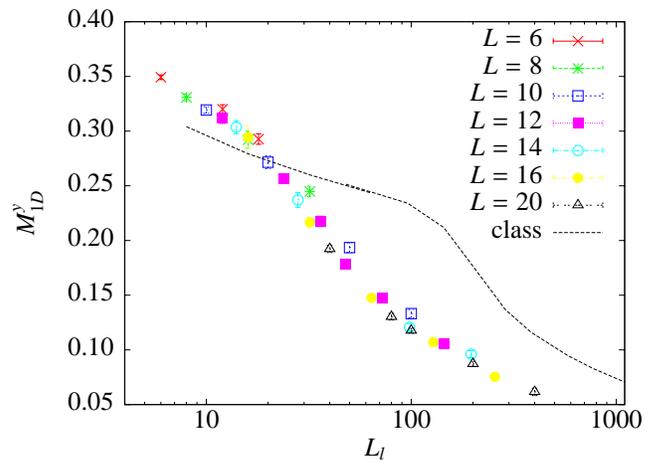}
\end{minipage}
\caption{\label{fig:QCM:M1D}(Color online) Analysis of the one-dimensional
  magnetization $M^y_{1D}$ for the quantum CM.  The result from the
  classical case is indicated by the line taken from
  Fig.~\ref{fig:MAG1D} (and divided by a factor $2$). The magnetic
  length scale is clearly much larger in the classical case. }
\end{figure}

We expect that SBC rectify and improve this value. Therefore, new QMC
simulations in the stochastic series expansion (SSE) framework using directed-loops
\cite{sandvik_operatorloops,syljuasen_dl} and PT updates were
performed implementing $S=1$ SBC. Otherwise our approach rests on that
presented in Ref.~\cite{wenzelQCMPRB} where concrete implementation
issues are discussed. A couple of simulations for lattice sizes
$L=10,12,16,20,24,28,32,42$ were performed and approximately $100\,000$
statistically independent samples of the order parameter $D$ were
taken in each case. The pseudocritical temperatures $T_{\max}(L)$ were obtained
from the peak in the variance of $D$ utilizing the quantum
generalization of the multihistogram reweighting idea
\cite{Troyer_multihist}. Figure \ref{fig:ffs} shows the pseudo\-critical
temperatures obtained and compares them to the old data utilizing
PBC. As expected, an evidently improved FSS
behavior is observed for the screw-periodic case. This is apparent
from the absolute move of $T_{\max}$ towards the true critical
temperature for small $L$ \emph{and} the much better power-law scaling
in terms of $1/L$. Indeed, the SBC data are fully consistent with
$\nu=1$ and a straight line fit to
\begin{equation}
T_{\max}(L) =T_\mathrm{c}+aL^{-1}\,
\end{equation}
yields our new estimate for the critical temperature as
\begin{equation}
T_\mathrm{c}=0.0585(3)
\end{equation}
with $\chi^2/\mathrm{d.o.f}=1.5$ using all lattice sizes studied.
Even using only the of smallest systems $L=12$ to $L=20$ the
extrapolation yields a consistent value of $T_\mathrm{c}=0.058(1)$, a
property which is of most practical relevance for studies aiming at
numerically verifying more qualitative effects (see, e.g., Ref.\ \cite{tanaka:256402}). Leaving $\nu$ as a free fit parameter as in
Eq.~\eqref{eqn:FSST} we obtain $T_\mathrm{c}=0.0586(8)$ and
$\nu=0.97(15)$ which is consistent with 2D Ising behavior.
Hence, although we have performed much less simulations \emph{and} on
smaller system sizes, we have obtained a much better and improved
result just by an adequate choice of the boundary conditions. The present
result does not agree within error bars with the previous estimate
$T_\mathrm{c}=0.055(1)$ because of the anomalous behavior which was
not accounted for in the ordinary Ising extrapolation (dashed line in
Fig.~\ref{fig:ffs}) with a $L^{-\omega}$ correction on periodic
lattices.
\begin{table}[b]
  \caption{\label{tab:Tcoverview}Previous and current estimates of the critical temperature of the DO transition exemplifying the (previous) difficulty of its extraction.}
\begin{tabular}{l c c c c}
\hline\hline
$T_\mathrm{c}$ & System sizes & Boundary cond. & Method & Ref. \\
\hline
$0.075(2)$  & $10-20$ & Periodic & Trotter QMC & \cite{tanaka:256402} \\
$0.055(1)$  & $10-96$ & Periodic & SSE + PT & \cite{wenzelQCMPRB} \\[0.2cm]
$0.058(1)$  & $12-20$ & Screw & SSE + PT & This work \\
$0.0585(3)$ & $12-42$ & Screw & SSE + PT & This work \\
\hline\hline
\end{tabular}
\end{table}
However, it appears that the effect of the magnetic length scale is not as
severe as in the classical case. This could be expected due to the
presence of quantum fluctuations. On the other hand the temperature
regime is lower which could in principle even stabilize the unwanted
order. In order to get an approximate estimate for the length scale
involved, we have finally analyzed the one-dimensional magnetization
also for the quantum case, where we restrict ourselves to measure
$M_{1D}^y$ (which corresponds to the quantization direction) along
$y$-loops. 
 Figure \ref{fig:QCM:M1D} shows the SSE estimates for $M_{1D}^y$ 
for various system sizes and screw parameters at $T=0.07$
(chosen for convenience because it is close to
  $T_\mathrm{c}$ but still in the region where PBC show unusual
  behavior). It verifies that quantum fluctuations reduce the overall
value of $M^y_{1D}$ and that they lead to a clear diminution of
pseudo-magnetic order at a scale corresponding to roughly $L\approx
50$ which is apparently smaller than in the classical case (line in
Fig.~\ref{fig:QCM:M1D}). Moreover, we also arrive at this
  conclusion by studying the behavior of the susceptibility (similar
  to that in Fig.~\ref{fig:screw_suscmp}) and find that the resonance
  is shifted to a smaller length scale in accordance with the findings
  just described. However, even such a moderate scale can still be a
formidable challenge to overcome for QMC without SBC.

In summary, the estimate of $T_\mathrm{c}$ for the quantum CM has seen
several steps of adjustments on a relatively short time scale as
summarized in Table~\ref{tab:Tcoverview}. The result of this section
provides the new benchmark estimate which should be useful for future
studies.

\section{\label{sec:concl}Summary and Conclusions}

Summarizing, we have revisited the directional-ordering transition in the classical and
quantum compass models employing two types of methodological advances.

In the classical case we were able to formulate a special
one-dimensional cluster update which in combination with Metropolis
and PT methods allowed to investigate much larger system sizes than
before. The following detailed comparison between the classical CM
with periodic boundary conditions and a fluctuating bond ensemble
showed that periodic boundary conditions scale much worse than known
so far.  Instead of the usual power law, anomalous scaling becomes
evident with a resonance and non-monotonic behavior in the
susceptibility and the Binder parameter at length scale of about
$L\approx 100-200$. In any typical MC simulation one would therefore
not be able to predict critical properties correctly when the
simulation is done with periodic boundary conditions. This resonance
is argued to be due to a magnetic correlation length which prohibits
the formation of domain walls at finite temperature on small
clusters. To counteract this problem we have proposed to employ
screw-periodic boundary conditions. We have shown that they are able
to remove scaling anomalies in the classical case almost completely.

This concept then proved to be a key step for simulations of the
quantum compass model where we were able to obtain a more accurate
estimate of the critical DO temperature based only on the change in
boundary conditions. On the physical side we have seen that the CM
represents a formidable challenge despite its simplicity --- even for
well settled numerical approaches. The right choice of boundary
conditions or topology is more essential for numerical studies of the
CM than for many other models.

Technically, it is clear that screw-periodic boundary conditions
should be used in future studies of various other aspects in the
quantum compass model.  Moreover, we regard SBC as a well suited and
general method which deserves more attention even in studies of other
systems. Via the screw parameter $S$ one may be able to tune or
minimize corrections to scaling. We are currently applying them to
further studies of the quantum phase transition in 2D dimerized
Heisenberg models (see, e.g., Ref.~\cite{wenzelPRL}).

\acknowledgements S.W. thanks the MPIPKS for support through its
visitors program. The numerical work of this paper was performed on
the GRAWP cluster at the Institute for Theoretical Physics of the
University of Leipzig, on the CALLISTO cluster at EPFL, and on the
JUROPA capability computer at NIC/JSC, Forschungszentrum J\"ulich
under Grant No.~HLZ12.

\bibliographystyle{apsrev}
\bibliography{literature}

\begin{thebibliography}{41}
\expandafter\ifx\csname natexlab\endcsname\relax\def\natexlab#1{#1}\fi
\expandafter\ifx\csname bibnamefont\endcsname\relax
  \def\bibnamefont#1{#1}\fi
\expandafter\ifx\csname bibfnamefont\endcsname\relax
  \def\bibfnamefont#1{#1}\fi
\expandafter\ifx\csname citenamefont\endcsname\relax
  \def\citenamefont#1{#1}\fi
\expandafter\ifx\csname url\endcsname\relax
  \def\url#1{\texttt{#1}}\fi
\expandafter\ifx\csname urlprefix\endcsname\relax\def\urlprefix{URL }\fi
\providecommand{\bibinfo}[2]{#2}
\providecommand{\eprint}[2][]{\url{#2}}

\bibitem[{\citenamefont{Kugel and Khomskii}(1982)}]{kugel82}
\bibinfo{author}{\bibfnamefont{K.}~\bibnamefont{Kugel}} \bibnamefont{and}
  \bibinfo{author}{\bibfnamefont{D.}~\bibnamefont{Khomskii}},
  \bibinfo{journal}{Sov. Phys. Usp.} \textbf{\bibinfo{volume}{25}},
  \bibinfo{pages}{231} (\bibinfo{year}{1982}).

\bibitem[{\citenamefont{Dou\c{c}ot et~al.}(2005)\citenamefont{Dou\c{c}ot,
  Feigel'man, Ioffe, and Ioselevich}}]{doucot:024505}
\bibinfo{author}{\bibfnamefont{B.}~\bibnamefont{Dou\c{c}ot}},
  \bibinfo{author}{\bibfnamefont{M.V.}~\bibnamefont{Feigel'man}},
  \bibinfo{author}{\bibfnamefont{L.B.}~\bibnamefont{Ioffe}}, \bibnamefont{and}
  \bibinfo{author}{\bibfnamefont{A.S.}~\bibnamefont{Ioselevich}},
  \bibinfo{journal}{Phys. Rev. B} \textbf{\bibinfo{volume}{71}},
  \bibinfo{eid}{024505} (\bibinfo{year}{2005}).

\bibitem[{\citenamefont{Milman et~al.}(2007)\citenamefont{Milman, Maineult,
  Guibal, Guidoni, Dou\c{c}ot, Ioffe, and Coudreau}}]{milman:020503}
\bibinfo{author}{\bibfnamefont{P.}~\bibnamefont{Milman}},
  \bibinfo{author}{\bibfnamefont{W.}~\bibnamefont{Maineult}},
  \bibinfo{author}{\bibfnamefont{S.}~\bibnamefont{Guibal}},
  \bibinfo{author}{\bibfnamefont{L.}~\bibnamefont{Guidoni}},
  \bibinfo{author}{\bibfnamefont{B.}~\bibnamefont{Dou\c{c}ot}},
  \bibinfo{author}{\bibfnamefont{L.}~\bibnamefont{Ioffe}}, \bibnamefont{and}
  \bibinfo{author}{\bibfnamefont{T.}~\bibnamefont{Coudreau}},
  \bibinfo{journal}{Phys. Rev. Lett.} \textbf{\bibinfo{volume}{99}},
  \bibinfo{eid}{020503} (\bibinfo{year}{2007}).

\bibitem[{\citenamefont{Gladchenko et~al.}(2009)\citenamefont{Gladchenko,
  Olaya, Dupont-Ferrier, Dou\c{c}ot, Ioffe, and Gershenson}}]{gladchenko-2008}
\bibinfo{author}{\bibfnamefont{S.}~\bibnamefont{Gladchenko}},
  \bibinfo{author}{\bibfnamefont{D.}~\bibnamefont{Olaya}},
  \bibinfo{author}{\bibfnamefont{E.}~\bibnamefont{Dupont-Ferrier}},
  \bibinfo{author}{\bibfnamefont{B.}~\bibnamefont{Dou\c{c}ot}},
  \bibinfo{author}{\bibfnamefont{L.}~\bibnamefont{Ioffe}}, \bibnamefont{and}
  \bibinfo{author}{\bibfnamefont{M.}~\bibnamefont{Gershenson}},
  \bibinfo{journal}{Nature Physics} \textbf{\bibinfo{volume}{5}},
  \bibinfo{pages}{48} (\bibinfo{year}{2009}).

\bibitem[{\citenamefont{Jackeli and Khaliullin}(2009)}]{jackeli-2008}
\bibinfo{author}{\bibfnamefont{G.}~\bibnamefont{Jackeli}} \bibnamefont{and}
  \bibinfo{author}{\bibfnamefont{G.}~\bibnamefont{Khaliullin}},
  \bibinfo{journal}{Phys. Rev. Lett.} \textbf{\bibinfo{volume}{102}},
  \bibinfo{pages}{017205} (\bibinfo{year}{2009}).

\bibitem[{\citenamefont{Dorier et~al.}(2005)\citenamefont{Dorier, Becca, and
  Mila}}]{dorier:024448}
\bibinfo{author}{\bibfnamefont{J.}~\bibnamefont{Dorier}},
  \bibinfo{author}{\bibfnamefont{F.}~\bibnamefont{Becca}}, \bibnamefont{and}
  \bibinfo{author}{\bibfnamefont{F.}~\bibnamefont{Mila}},
  \bibinfo{journal}{Phys.~Rev. B} \textbf{\bibinfo{volume}{72}},
  \bibinfo{eid}{024448} (\bibinfo{year}{2005}).

\bibitem[{\citenamefont{Mishra et~al.}(2004)\citenamefont{Mishra, Ma, Zhang,
  Guertler, Tang, and Wan}}]{mishra:207201}
\bibinfo{author}{\bibfnamefont{A.}~\bibnamefont{Mishra}},
  \bibinfo{author}{\bibfnamefont{M.}~\bibnamefont{Ma}},
  \bibinfo{author}{\bibfnamefont{F.-C.} \bibnamefont{Zhang}},
  \bibinfo{author}{\bibfnamefont{S.}~\bibnamefont{Guertler}},
  \bibinfo{author}{\bibfnamefont{L.-H.} \bibnamefont{Tang}}, \bibnamefont{and}
  \bibinfo{author}{\bibfnamefont{S.}~\bibnamefont{Wan}},
  \bibinfo{journal}{Phys. Rev.~Lett.} \textbf{\bibinfo{volume}{93}},
  \bibinfo{eid}{207201} (\bibinfo{year}{2004}).

\bibitem[{\citenamefont{Tanaka and Ishihara}(2007)}]{tanaka:256402}
\bibinfo{author}{\bibfnamefont{T.}~\bibnamefont{Tanaka}} \bibnamefont{and}
  \bibinfo{author}{\bibfnamefont{S.}~\bibnamefont{Ishihara}},
  \bibinfo{journal}{Phys. Rev. Lett.} \textbf{\bibinfo{volume}{98}},
  \bibinfo{eid}{256402} (\bibinfo{year}{2007}).

\bibitem[{\citenamefont{Wenzel and Janke}(2008)}]{wenzelQCMPRB}
\bibinfo{author}{\bibfnamefont{S.}~\bibnamefont{Wenzel}} \bibnamefont{and}
  \bibinfo{author}{\bibfnamefont{W.}~\bibnamefont{Janke}},
  \bibinfo{journal}{Phys. Rev. B} \textbf{\bibinfo{volume}{78}},
  \bibinfo{pages}{064402} (\bibinfo{year}{2008}).

\bibitem[{\citenamefont{Brzezicki et~al.}(2007)\citenamefont{Brzezicki,
  Dziarmaga, and Ole\'{s}}}]{CM1DBrzezicki}
\bibinfo{author}{\bibfnamefont{W.}~\bibnamefont{Brzezicki}},
  \bibinfo{author}{\bibfnamefont{J.}~\bibnamefont{Dziarmaga}},
  \bibnamefont{and} \bibinfo{author}{\bibfnamefont{A.M.}~\bibnamefont{Ole\'{s}}},
  \bibinfo{journal}{Phys. Rev. B} \textbf{\bibinfo{volume}{75}},
  \bibinfo{eid}{134415} (\bibinfo{year}{2007}).

\bibitem[{\citenamefont{Eriksson and Johannesson}(2009)}]{eriksson:224424}
\bibinfo{author}{\bibfnamefont{E.}~\bibnamefont{Eriksson}} \bibnamefont{and}
  \bibinfo{author}{\bibfnamefont{H.}~\bibnamefont{Johannesson}},
  \bibinfo{journal}{Phys. Rev. B} \textbf{\bibinfo{volume}{79}},
  \bibinfo{pages}{224424} (\bibinfo{year}{2009}).

\bibitem[{\citenamefont{Sun et~al.}(2009)\citenamefont{Sun, Zhang, and
  Chen}}]{CM1DSun}
\bibinfo{author}{\bibfnamefont{K.-W.} \bibnamefont{Sun}},
  \bibinfo{author}{\bibfnamefont{Y.-Y.} \bibnamefont{Zhang}}, \bibnamefont{and}
  \bibinfo{author}{\bibfnamefont{Q.-H.} \bibnamefont{Chen}},
  \bibinfo{journal}{Phys.\ Rev.\ B} \textbf{\bibinfo{volume}{79}},
  \bibinfo{pages}{104429} (\bibinfo{year}{2009}).

\bibitem[{\citenamefont{{Scarola} et~al.}(2009)\citenamefont{{Scarola},
  {Whaley}, and {Troyer}}}]{scarola:CM}
\bibinfo{author}{\bibfnamefont{V.W.}~\bibnamefont{{Scarola}}},
  \bibinfo{author}{\bibfnamefont{K.B.}~\bibnamefont{{Whaley}}}, \bibnamefont{and}
  \bibinfo{author}{\bibfnamefont{M.}~\bibnamefont{{Troyer}}},
  \bibinfo{journal}{Phys. Rev. B} \textbf{\bibinfo{volume}{79}},
  \bibinfo{pages}{085113} (\bibinfo{year}{2009}).

\bibitem[{\citenamefont{Wenzel and Janke}(2009)}]{wenzelPOM}
\bibinfo{author}{\bibfnamefont{S.}~\bibnamefont{Wenzel}} \bibnamefont{and}
  \bibinfo{author}{\bibfnamefont{W.}~\bibnamefont{Janke}},
  \bibinfo{journal}{Phys. Rev. B} \textbf{\bibinfo{volume}{80}},
  \bibinfo{pages}{054403} (\bibinfo{year}{2009}).

\bibitem[{\citenamefont{Xu and Moore}(2004)}]{PhysRevLett.93.047003}
\bibinfo{author}{\bibfnamefont{C.}~\bibnamefont{Xu}} \bibnamefont{and}
  \bibinfo{author}{\bibfnamefont{J.E.}~\bibnamefont{Moore}},
  \bibinfo{journal}{Phys. Rev. Lett.} \textbf{\bibinfo{volume}{93}},
  \bibinfo{pages}{047003} (\bibinfo{year}{2004}).

\bibitem[{\citenamefont{Nussinov and Fradkin}(2005)}]{nussinovpip}
\bibinfo{author}{\bibfnamefont{Z.}~\bibnamefont{Nussinov}} \bibnamefont{and}
  \bibinfo{author}{\bibfnamefont{E.}~\bibnamefont{Fradkin}},
  \bibinfo{journal}{Phys. Rev. B} \textbf{\bibinfo{volume}{71}},
  \bibinfo{pages}{195120} (\bibinfo{year}{2005}).

\bibitem[{\citenamefont{Batista and Nussinov}(2005)}]{Batista_dimred2005}
\bibinfo{author}{\bibfnamefont{C.D.}~\bibnamefont{Batista}} \bibnamefont{and}
  \bibinfo{author}{\bibfnamefont{Z.}~\bibnamefont{Nussinov}},
  \bibinfo{journal}{Phys. Rev. B} \textbf{\bibinfo{volume}{72}},
  \bibinfo{pages}{045137} (\bibinfo{year}{2005}).

\bibitem[{\citenamefont{Vidal et~al.}(2009)\citenamefont{Vidal, Thomale,
  Schmidt, and Dusuel}}]{KaiToricTransverse}
\bibinfo{author}{\bibfnamefont{J.}~\bibnamefont{Vidal}},
  \bibinfo{author}{\bibfnamefont{R.}~\bibnamefont{Thomale}},
  \bibinfo{author}{\bibfnamefont{K.~P.} \bibnamefont{Schmidt}},
  \bibnamefont{and} \bibinfo{author}{\bibfnamefont{S.}~\bibnamefont{Dusuel}},
  \bibinfo{journal}{Phys. Rev. B} \textbf{\bibinfo{volume}{80}},
  \bibinfo{pages}{081104(R)} (\bibinfo{year}{2009}).

\bibitem[{\citenamefont{Kitaev}(2003)}]{kitaev-toriccode}
\bibinfo{author}{\bibfnamefont{A.~Y.} \bibnamefont{Kitaev}},
  \bibinfo{journal}{Ann. Phys.} \textbf{\bibinfo{volume}{303}},
  \bibinfo{pages}{2} (\bibinfo{year}{2003}).

\bibitem[{\citenamefont{Chen et~al.}(2007)\citenamefont{Chen, Fang, Hu, and
  Yao}}]{CMfirstorder}
\bibinfo{author}{\bibfnamefont{H.-D.} \bibnamefont{Chen}},
  \bibinfo{author}{\bibfnamefont{C.}~\bibnamefont{Fang}},
  \bibinfo{author}{\bibfnamefont{J.}~\bibnamefont{Hu}}, \bibnamefont{and}
  \bibinfo{author}{\bibfnamefont{H.}~\bibnamefont{Yao}},
  \bibinfo{journal}{Phys. Rev. B} \textbf{\bibinfo{volume}{75}},
  \bibinfo{pages}{144401} (\bibinfo{year}{2007}).

\bibitem[{\citenamefont{Or\'us et~al.}(2009)\citenamefont{Or\'us, Doherty, and
  Vidal}}]{CMorusfirstorder}
\bibinfo{author}{\bibfnamefont{R.}~\bibnamefont{Or\'us}},
  \bibinfo{author}{\bibfnamefont{A.C.}~\bibnamefont{Doherty}}, \bibnamefont{and}
  \bibinfo{author}{\bibfnamefont{G.}~\bibnamefont{Vidal}},
  \bibinfo{journal}{Phys. Rev. Lett.} \textbf{\bibinfo{volume}{102}},
  \bibinfo{pages}{077203} (\bibinfo{year}{2009}).

\bibitem[{\citenamefont{Jordan et~al.}(2008)\citenamefont{Jordan, Or\'{u}s,
  Vidal, Verstraete, and Cirac}}]{iPEPS}
\bibinfo{author}{\bibfnamefont{J.}~\bibnamefont{Jordan}},
  \bibinfo{author}{\bibfnamefont{R.}~\bibnamefont{Or\'{u}s}},
  \bibinfo{author}{\bibfnamefont{G.}~\bibnamefont{Vidal}},
  \bibinfo{author}{\bibfnamefont{F.}~\bibnamefont{Verstraete}},
  \bibnamefont{and} \bibinfo{author}{\bibfnamefont{J.~I.} \bibnamefont{Cirac}},
  \bibinfo{journal}{Phys. Rev. Lett.} \textbf{\bibinfo{volume}{101}},
  \bibinfo{pages}{250602} (\bibinfo{year}{2008}).

\bibitem{Cincio}
L. Cincio, J. Dziarmaga, and A. M. Ole\'{s}, arXiv:1001.5457 (2010).

\bibitem{MERA}
G. Vidal, Phys. Rev. Lett. \textbf{101}, 110501 (2008).

\bibitem[{\citenamefont{Geyer and Thompson}(1995)}]{geyerPT}
\bibinfo{author}{\bibfnamefont{C.}~\bibnamefont{Geyer}} \bibnamefont{and}
  \bibinfo{author}{\bibfnamefont{E.}~\bibnamefont{Thompson}},
  \bibinfo{journal}{J. Am. Stat. Assoc.} \textbf{\bibinfo{volume}{90}},
  \bibinfo{pages}{909} (\bibinfo{year}{1995}).

\bibitem[{\citenamefont{Hukushima and Nemoto}(1996)}]{hukushimaPT}
\bibinfo{author}{\bibfnamefont{K.}~\bibnamefont{Hukushima}} \bibnamefont{and}
  \bibinfo{author}{\bibfnamefont{K.}~\bibnamefont{Nemoto}},
  \bibinfo{journal}{J. Phys. Soc. Jpn.} \textbf{\bibinfo{volume}{65}},
  \bibinfo{pages}{1604} (\bibinfo{year}{1996}).

\bibitem[{\citenamefont{Bittner et~al.}(2008)\citenamefont{Bittner,
  Nu\ss{}baumer, and Janke}}]{BittnerPT}
\bibinfo{author}{\bibfnamefont{E.}~\bibnamefont{Bittner}},
  \bibinfo{author}{\bibfnamefont{A.}~\bibnamefont{Nu\ss{}baumer}},
  \bibnamefont{and} \bibinfo{author}{\bibfnamefont{W.}~\bibnamefont{Janke}},
  \bibinfo{journal}{Phys. Rev. Lett.} \textbf{\bibinfo{volume}{101}},
  \bibinfo{pages}{130603} (\bibinfo{year}{2008}).

\bibitem[{\citenamefont{Pelissetto and Vicari}(2002)}]{vicarireview}
\bibinfo{author}{\bibfnamefont{A.}~\bibnamefont{Pelissetto}} \bibnamefont{and}
  \bibinfo{author}{\bibfnamefont{E.}~\bibnamefont{Vicari}},
  \bibinfo{journal}{Phys. Rep.} \textbf{\bibinfo{volume}{368}},
  \bibinfo{pages}{549} (\bibinfo{year}{2002}).

\bibitem[{\citenamefont{{D.P.~Landau} and {K.~Binder}}(2000)}]{LandauMCBook}
\bibinfo{author}{\bibnamefont{{D.P.~Landau}}} \bibnamefont{and}
  \bibinfo{author}{\bibnamefont{{K.~Binder}}}, \emph{\bibinfo{title}{A Guide to
  Monte Carlo Simulations in Statistical Physics}}
  (\bibinfo{publisher}{Cambridge University Press},
  \bibinfo{address}{Cambridge}, \bibinfo{year}{2000}), \bibinfo{edition}{1st}
  ed.

\bibitem[{\citenamefont{Janke}(2008)}]{JankeMCGreifswald}
\bibinfo{author}{\bibfnamefont{W.}~\bibnamefont{Janke}},
  \bibinfo{journal}{Lect. Notes Phys.} \textbf{\bibinfo{volume}{739}},
  \bibinfo{pages}{79} (\bibinfo{year}{2008}).

\bibitem[{\citenamefont{{A.M.~Ferrenberg} and
  {R.H.~Swendsen}}(1989)}]{ferrenberg:multi}
\bibinfo{author}{\bibnamefont{{A.M.~Ferrenberg}}} \bibnamefont{and}
  \bibinfo{author}{\bibnamefont{{R.H.~Swendsen}}}, \bibinfo{journal}{Phys. Rev.
  Lett.} \textbf{\bibinfo{volume}{63}}, \bibinfo{pages}{1195}
  (\bibinfo{year}{1989}).

\bibitem[{\citenamefont{Efron}(1982)}]{efron}
\bibinfo{author}{\bibfnamefont{B.}~\bibnamefont{Efron}},
  \emph{\bibinfo{title}{The Jackknife, the Bootstrap, and other Resampling
  Plans}} (\bibinfo{publisher}{Society for Industriell and Applied Mathematics
  [SIAM]}, \bibinfo{address}{Philadelphia}, \bibinfo{year}{1982}).

\bibitem[{\citenamefont{Wolff}(1989)}]{WolffCluster}
\bibinfo{author}{\bibfnamefont{U.}~\bibnamefont{Wolff}},
  \bibinfo{journal}{Phys. Rev. Lett.} \textbf{\bibinfo{volume}{62}},
  \bibinfo{pages}{361} (\bibinfo{year}{1989}).

\bibitem[{\citenamefont{{M.E.J.~Newman} and
  {G.T.~Barkema}}(1999)}]{BarkemaMCBook}
\bibinfo{author}{\bibnamefont{{M.E.J.~Newman}}} \bibnamefont{and}
  \bibinfo{author}{\bibnamefont{{G.T.~Barkema}}}, \emph{\bibinfo{title}{Monte
  Carlo Methods in Statistical Physics}} (\bibinfo{publisher}{Oxford University
  Press}, \bibinfo{address}{Oxford}, \bibinfo{year}{1999}),
  \bibinfo{edition}{1st} ed.

\bibitem[{\citenamefont{Kitatani and Oguchi}(1992)}]{Ising_screwbc}
\bibinfo{author}{\bibfnamefont{H.}~\bibnamefont{Kitatani}} \bibnamefont{and}
  \bibinfo{author}{\bibfnamefont{T.}~\bibnamefont{Oguchi}},
  \bibinfo{journal}{J. Phys. Soc. Jpn.} \textbf{\bibinfo{volume}{61}},
  \bibinfo{pages}{1598} (\bibinfo{year}{1992}).

\bibitem[{\citenamefont{Bittner et~al.}(2009)\citenamefont{Bittner,
  Nu\ss{}baumer, and Janke}}]{BittnerscrewBC}
\bibinfo{author}{\bibfnamefont{E.}~\bibnamefont{Bittner}},
  \bibinfo{author}{\bibfnamefont{A.}~\bibnamefont{Nu\ss{}baumer}},
  \bibnamefont{and} \bibinfo{author}{\bibfnamefont{W.}~\bibnamefont{Janke}},
  \bibinfo{journal}{Nucl. Phys. B} \textbf{\bibinfo{volume}{820}},
  \bibinfo{pages}{694} (\bibinfo{year}{2009}).

\bibitem[{\citenamefont{Kamieniarz and Bl\"ote}(1993)}]{Kamieniarz}
\bibinfo{author}{\bibfnamefont{G.}~\bibnamefont{Kamieniarz}} \bibnamefont{and}
  \bibinfo{author}{\bibfnamefont{H.}~\bibnamefont{Bl\"ote}},
  \bibinfo{journal}{J. Phys. A: Math. and Gen.} \textbf{\bibinfo{volume}{26}},
  \bibinfo{pages}{201} (\bibinfo{year}{1993}).

\bibitem[{\citenamefont{Salas and Sokal}(2000)}]{SalasBinderIsing}
\bibinfo{author}{\bibfnamefont{J.}~\bibnamefont{Salas}} \bibnamefont{and}
  \bibinfo{author}{\bibfnamefont{A.~D.} \bibnamefont{Sokal}},
  \bibinfo{journal}{J. Stat. Phys.} \textbf{\bibinfo{volume}{98}},
  \bibinfo{pages}{551} (\bibinfo{year}{2000}).

\bibitem[{\citenamefont{Selke}(2006)}]{SelkeBinderBC}
\bibinfo{author}{\bibfnamefont{W.}~\bibnamefont{Selke}}, \bibinfo{journal}{Eur.
  Phys. J. B} \textbf{\bibinfo{volume}{51}}, \bibinfo{pages}{223}
  (\bibinfo{year}{2006}).

\bibitem[{\citenamefont{Sandvik}(1999)}]{sandvik_operatorloops}
\bibinfo{author}{\bibfnamefont{A.W.}~\bibnamefont{Sandvik}},
  \bibinfo{journal}{Phys. Rev. B} \textbf{\bibinfo{volume}{59}},
  \bibinfo{pages}{R14157} (\bibinfo{year}{1999}).

\bibitem[{\citenamefont{Sylju\aa{}sen and Sandvik}(2002)}]{syljuasen_dl}
\bibinfo{author}{\bibfnamefont{O.F.}~\bibnamefont{Sylju\aa{}sen}}
  \bibnamefont{and} \bibinfo{author}{\bibfnamefont{A.W.}~\bibnamefont{Sandvik}},
  \bibinfo{journal}{Phys. Rev. E} \textbf{\bibinfo{volume}{66}},
  \bibinfo{pages}{046701} (\bibinfo{year}{2002}).

\bibitem[{\citenamefont{Troyer et~al.}(2004)\citenamefont{Troyer, Wessel, and
  Alet}}]{Troyer_multihist}
\bibinfo{author}{\bibfnamefont{M.}~\bibnamefont{Troyer}},
  \bibinfo{author}{\bibfnamefont{S.}~\bibnamefont{Wessel}}, \bibnamefont{and}
  \bibinfo{author}{\bibfnamefont{F.}~\bibnamefont{Alet}},
  \bibinfo{journal}{Braz. J. Phys.} \textbf{\bibinfo{volume}{34}},
  \bibinfo{pages}{377} (\bibinfo{year}{2004}).

\bibitem[{\citenamefont{Wenzel et~al.}(2008)\citenamefont{Wenzel, Bogacz, and
  Janke}}]{wenzelPRL}
\bibinfo{author}{\bibfnamefont{S.}~\bibnamefont{Wenzel}},
  \bibinfo{author}{\bibfnamefont{L.}~\bibnamefont{Bogacz}}, \bibnamefont{and}
  \bibinfo{author}{\bibfnamefont{W.}~\bibnamefont{Janke}},
  \bibinfo{journal}{Phys. Rev. Lett.} \textbf{\bibinfo{volume}{101}},
  \bibinfo{pages}{127202} (\bibinfo{year}{2008}).

\end{thebibliography}
\end{document}